\documentclass[12pt]{article}
\usepackage{graphicx}
\def\beq{\begin{equation}}
\def\eeq{\end{equation}}
\def\bea{\begin{eqnarray}}
\def\eea{\end{eqnarray}}
\def\nn{\nonumber}
\def\sss{\scriptscriptstyle}
\def\bd{B_d^0}
\def\bdbar{{\bar B}_d^0}

\def\barp{{\raise.35ex\hbox
{${\sss (}$}}---{\raise.35ex\hbox{${\sss )}$}}}
\def\bdbarp{\hbox{$B_d$\kern-1.4em\raise1.4ex\hbox{\barp}}}
\def\bsbarp{\hbox{$B_s$\kern-1.4em\raise1.4ex\hbox{\barp}}}
\def\ks{K_{\sss S}}

\def\barpk{{\raise.35ex\hbox
{${\sss (}$}}--{\raise.35ex\hbox{${\sss )}$}}}
\def\kbarp{\hbox{$K$\kern-0.9em\raise1.4ex\hbox{\barpk}}}
\def\bbarp{\hbox{$B$\kern-0.9em\raise1.4ex\hbox{\barpk}}}
\def\roughly#1{\mathrel{\raise.3ex\hbox
{$#1$\kern-.75em\lower1ex\hbox{$\sim$}}}}

\def\epjc#1#2#3{{\it Eur.\ Phys.\ J.}\ {\bf C#1}, #3, #2}

\def\plb#1#2#3{{\it Phys.\ Lett.} {\bf #1B}, #3, #2}
\def\prd#1#2#3{{\it Phys.\ Rev.} {\bf D#1}, #3, #2}
\def\newprd#1#2#3{{\it Phys.\ Rev.} {\bf D#1}: #3, #2}
\def\prl#1#2#3{{\it Phys.\ Rev.\ Lett.} {\bf #1}, #3, #2}

\newread\epsffilein 
\newif\ifepsffileok 
\newif\ifepsfbbfound 
\newif\ifepsfverbose 
\newdimen\epsfxsize 
\newdimen\epsfysize 
\newdimen\epsftsize 
\newdimen\epsfrsize 
\newdimen\epsftmp 
\newdimen\pspoints 
\def\epsfbox#1{\global\def\epsfllx{72}\global\def\epsflly{72}%
 \global\def\epsfurx{540}\global\def\epsfury{720}%
 \def\lbracket{[}\def\testit{#1}\ifx\testit\lbracket
 \let\next=\epsfgetlitbb\else\let\next=\epsfnormal\fi\next{#1}}%
\def\epsfgetlitbb#1#2 #3 #4 #5]#6{\epsfgrab #2 #3 #4 #5 .\\%
 \epsfsetgraph{#6}}%
\def\epsfnormal#1{\epsfgetbb{#1}\epsfsetgraph{#1}}%
\def\epsfgetbb#1{%
\openin\epsffilein=#1
\ifeof\epsffilein\errmessage{I couldn't open #1, will ignore it}\else
 {\epsffileoktrue \chardef\other=12
 \def\do##1{\catcode`##1=\other}\dospecials \catcode`\ =10
 \loop
 \read\epsffilein to \epsffileline
 \ifeof\epsffilein\epsffileokfalse\else
 \expandafter\epsfaux\epsffileline:. \\%
 \fi
 \ifepsffileok\repeat
 \ifepsfbbfound\else
 \ifepsfverbose\message{No bounding box comment in #1; using defaults}\fi\fi
 }\closein\epsffilein\fi}%
\def\epsfclipstring{}
\def\epsfsetgraph#1{%
 \epsfrsize=\epsfury\pspoints
 \advance\epsfrsize by-\epsflly\pspoints
 \epsftsize=\epsfurx\pspoints
 \advance\epsftsize by-\epsfllx\pspoints
 \epsfxsize\epsfsize\epsftsize\epsfrsize
 \ifnum\epsfxsize=0 \ifnum\epsfysize=0
 \epsfxsize=\epsftsize \epsfysize=\epsfrsize
 \epsfrsize=0pt
 \else\epsftmp=\epsftsize \divide\epsftmp\epsfrsize
 \epsfxsize=\epsfysize \multiply\epsfxsize\epsftmp
 \multiply\epsftmp\epsfrsize \advance\epsftsize-\epsftmp
 \epsftmp=\epsfysize
 \loop \advance\epsftsize\epsftsize \divide\epsftmp 2
 \ifnum\epsftmp>0
 \ifnum\epsftsize<\epsfrsize\else
 \advance\epsftsize-\epsfrsize \advance\epsfxsize\epsftmp \fi
 \repeat
 \epsfrsize=0pt
 \fi
 \else \ifnum\epsfysize=0
 \epsftmp=\epsfrsize \divide\epsftmp\epsftsize
 \epsfysize=\epsfxsize \multiply\epsfysize\epsftmp
 \multiply\epsftmp\epsftsize \advance\epsfrsize-\epsftmp
 \epsftmp=\epsfxsize
 \loop \advance\epsfrsize\epsfrsize \divide\epsftmp 2
 \ifnum\epsftmp>0
 \ifnum\epsfrsize<\epsftsize\else
 \advance\epsfrsize-\epsftsize \advance\epsfysize\epsftmp \fi
 \repeat
 \epsfrsize=0pt
 \else
 \epsfrsize=\epsfysize
 \fi
 \fi
 \ifepsfverbose\message{#1: width=\the\epsfxsize, height=\the\epsfysize}\fi
 \epsftmp=10\epsfxsize \divide\epsftmp\pspoints
 \vbox to\epsfysize{\vfil\hbox to\epsfxsize{%
 \ifnum\epsfrsize=0\relax
 \includegraphics{#1}%
 \else
 \epsfrsize=10\epsfysize \divide\epsfrsize\pspoints
 \includegraphics{#1}%
 \fi
 \hfil}}%
\global\epsfxsize=0pt\global\epsfysize=0pt}%
 {\catcode`\%=12 \global\let\epsfpercent=
\long\def\epsfaux#1#2:#3\\{\ifx#1\epsfpercent
 \def\testit{#2}\ifx\testit\epsfbblit
 \epsfgrab #3 . . . \\%
 \epsffileokfalse
 \global\epsfbbfoundtrue
 \fi\else\ifx#1\par\else\epsffileokfalse\fi\fi}%
\def\epsfempty{}%
\def\epsfgrab #1 #2 #3 #4 #5\\{%
\global\def\epsfllx{#1}\ifx\epsfllx\epsfempty
 \epsfgrab #2 #3 #4 #5 .\\\else
 \global\def\epsflly{#2}%
 \global\def\epsfurx{#3}\global\def\epsfury{#4}\fi}%
\def\epsfsize#1#2{\epsfxsize}
\pagestyle{plain}

\begin{document}

\begin{flushright}  
UdeM-GPP-TH-03-109 \\
\end{flushright}

\begin{center}
\bigskip
{\Large \bf Probing New Physics via an Angular Analysis \\ of $B \to
V_1 V_2$ Decays}
\end{center}

\begin{center}
{\large David London $^{a,}$\footnote{london@lps.umontreal.ca}, Nita
Sinha $^{b,}$\footnote{nita@imsc.res.in} and Rahul Sinha
$^{b,}$\footnote{sinha@imsc.res.in}}
\end{center}

\begin{flushleft}
~~~~~~~~~~~$a$: {\it Laboratoire Ren\'e J.-A. L\'evesque, 
Universit\'e de Montr\'eal,}\\
~~~~~~~~~~~~~~~{\it C.P. 6128, succ. centre-ville, Montr\'eal, QC,
Canada H3C 3J7} \\
~~~~~~~~~~~$b$: {\it Institute of Mathematical Sciences, C. I. T
 Campus,}\\
~~~~~~~~~~~~~~~{\it Taramani, Chennai 600 113, India}
\end{flushleft}

\begin{center} 
\bigskip (\today)
\vskip0.5cm
{\Large Abstract\\}
\vskip3truemm
\parbox[t]{\textwidth} {We show that an angular analysis of $B \to V_1
V_2$ decays yields numerous tests for new physics (NP) in the decay
amplitudes. Many of these NP observables are nonzero even if the
strong phase differences vanish. For certain observables, neither
time-dependent measurements nor tagging is necessary. Should a signal
for new physics be found, one can place a lower limit on the size of
the NP parameters, as well as on their effect on the measurement of
the phase of $B^0$--${\bar B}^0$ mixing.}
\end{center}

\thispagestyle{empty}
\newpage
\setcounter{page}{1}
\baselineskip=14pt

CP violation in the standard model (SM) is due to the presence of a
nonzero complex phase in the Cabibbo-Kobayashi-Maskawa (CKM) quark
mixing matrix. This explanation can be tested by measuring
CP-violating rate asymmetries in $B$ decays, and extracting $\alpha$,
$\beta$ and $\gamma$, the three interior angles of the unitarity
triangle \cite{CPreview}. If the measured values of these angles are
inconsistent with the predictions of the SM, this will indicate the
presence of new physics (NP).

The most promising modes for measuring the CP phases are those that
are dominated by a single decay amplitude. In this case, the
weak-phase information can be extracted cleanly, i.e.\ with no
hadronic uncertainties. An example of such a decay is the so-called
``gold-plated'' mode $\bd(t)\to J/\psi \ks$, which is used to probe
$\beta$~\footnote{In fact, there are two weak amplitudes that can
contribute to $\bd(t)\to J/\psi \ks$: the tree amplitude and the $b\to
s$ penguin amplitude. However, the weak phases of these two amplitudes
are equal (they vanish in the Wolfenstein parametrization of the CKM
matrix \cite{Wolfenstein}), so that there is effectively only a single
weak amplitude contributing to $\bd\to J/\psi \ks$. Thus, the
extraction of the CP phase $\beta$ from this decay mode is extremely
clean.}. Note that the decay $\bd(t) \to J/\psi K^*$ is equally
gold-plated. The only difference, in comparison to $\bd(t)\to
J/\psi\ks$, is that here the final state consists of two vector
particles. In this case, one has to do an angular analysis to separate
out the CP-even and CP-odd components \cite{CPreview}. Each component
can then be treated separately, and $\beta$ can be obtained cleanly.

Suppose now that there is new physics. How does this affect the above
analysis? If the NP affects $\bd$--$\bdbar$ mixing only, the above
analysis is unchanged, except that the measured value of $\beta$ is
not the true SM value, but rather one that has been shifted by a
new-physics phase. On the other hand, if the NP affects the decay
amplitude \cite{GrossWorah}, then the extraction of $\beta$ is no
longer clean -- it may be contaminated by hadronic uncertainties. It
is this situation that interests us in this paper.

New physics can affect the decay amplitude either at loop level (i.e.\
in the $b\to s$ penguin amplitude) or at tree level. Examples of such
new-physics models include non-minimal supersymmetric models and
models with $Z$-mediated flavor-changing neutral currents
\cite{newphysics}. In all cases, if the new contributions have a
different weak phase than that of the SM amplitude, then the measured
value of $\beta$, $\beta^{meas}$, no longer corresponds to the phase
of $\bd$--$\bdbar$ mixing, $\beta^{mix}$. (Note that $\beta^{mix}$
could include NP contributions to the mixing.)

If new-physics contributions to the decay amplitude are present, how
will we see them? One way is to note that, to a good approximation,
$\beta$ can also be obtained in the SM from $\bd(t) \to \phi \ks$ and
$\bd(t) \to \eta' \ks$ \cite{LonSoni}. Indeed, at present there
appears to be a discrepancy between the value of $\beta$ extracted
from $\bd(t)\to J/\psi \ks$ and that obtained from $\bd(t) \to \phi
\ks$ \cite{phiKs}. Should this difference remain as more data is
taken, it would provide indirect evidence for a NP amplitude in $B \to
\phi K$.

Still, even in this case, it would be preferable to have {\it direct}
evidence for this new amplitude. Furthermore, if present, we would
like to obtain information about it (magnitude, weak and strong
phases). It is therefore important to have as many independent tests
as possible for NP. One possibility is to search for direct CP
violation. However, direct CP asymmetries vanish if the strong phase
difference between the SM and NP amplitudes is zero, which may well be
the case in $B$ decays.  (It has been argued that all strong phases in
$B$ decays should be quite small, due to the fact that the $b$-quark
is rather heavy.)

In this Letter we show that an angular analysis of $B$-meson decays to
two vector mesons, such as $\bd(t) \to J/\psi K^*$ or $\phi K^*$, can
provide many signals for NP, including several that are nonzero even
if the strong phase differences vanish. Furthermore, if {\it any} NP
signal is found, this analysis allows one to place a lower bound on
the size of the NP amplitude, and on the difference $|\beta^{meas} -
\beta^{mix}|$. As we will see, the analysis can even be used within
the SM to analyze decays such as $\bd(t) \to D^{*+} D^{*-}$, from
which CP phases cannot be extracted cleanly due to penguin
``pollution.''

Consider a $B\to V_1 V_2$ decay for which a single weak decay
amplitude contributes within the SM. Suppose that there is a
new-physics amplitude, with a different weak phase, that contributes
to the decay. The decay amplitude for each of the three possible
helicity states may be generally written as
\begin{eqnarray}
A_\lambda \equiv Amp (B \to V_1V_2)_\lambda &=& a_\lambda e^{i
\delta_\lambda^a} + b_\lambda e^{i\phi} e^{i \delta_\lambda^b} ~,
\nn\\
{\bar A}_\lambda \equiv Amp ({\bar B} \to
{\overline{V}}_1 {\overline{V}}_2)_\lambda &=& a_\lambda e^{i
\delta_\lambda^a} + b_\lambda e^{-i\phi} e^{i \delta_\lambda^b} ~,
\label{amps}
\end{eqnarray}
where $a_\lambda$ and $b_\lambda$ represent the SM and NP amplitudes,
respectively, $\phi$ is the new-physics weak phase, the
$\delta_\lambda^{a,b}$ are the strong phases, and the helicity index
$\lambda$ takes the values $\left\{ 0,\|,\perp \right\}$. Using CPT
invariance, the full decay amplitudes can be written as
\begin{eqnarray}
{\cal A} &=& Amp (B\to V_1V_2) = A_0 g_0 + A_\| g_\| + i \, A_\perp
g_\perp~, \nn\\
{\bar{\cal A}} &=& Amp ({\bar B} \to {\overline{V}}_1
{\overline{V}}_2) = {\bar A}_0 g_0 + {\bar A}_\| g_\| - i \, {\bar
A}_\perp g_\perp~,
\label{fullamps}
\end{eqnarray}
where the $g_\lambda$ are the coefficients of the helicity amplitudes
written in the linear polarization basis. The $g_\lambda$ depend only
on the angles describing the kinematics \cite{glambda}. For $B = \bd$,
the above equations enable us to write the time-dependent decay rates
as
\beq
\label{decayrates}
\Gamma(\bdbarp(t) \to V_1V_2) = e^{-\Gamma t} \sum_{\lambda\leq\sigma}
\Bigl(\Lambda_{\lambda\sigma} \pm \Sigma_{\lambda\sigma}\cos(\Delta M
t) \mp \rho_{\lambda\sigma}\sin(\Delta M t)\Bigr) g_\lambda g_\sigma
~.
\eeq
Thus, by performing a time-dependent angular analysis of the decay
$\bd(t) \to V_1V_2$, one can measure 18 observables. These are:
{\setlength\arraycolsep{2pt}
\begin{eqnarray}
\Lambda_{\lambda\lambda}=\displaystyle
\frac{1}{2}(|A_\lambda|^2+|{\bar A}_\lambda|^2),&&
\Sigma_{\lambda\lambda}=\displaystyle
\frac{1}{2}(|A_\lambda|^2-|{\bar A}_\lambda|^2),\nn \\[1.ex]
\Lambda_{\perp i}= -\!{\rm Im}({ A}_\perp { A}_i^* \!-\! {\bar
A}_\perp {{\bar A}_i}^* ),
&&\Lambda_{\| 0}= {\rm Re}(A_\| A_0^*\! +\! {\bar A}_\| {{\bar A}_0}^*
), \nn \\[1.ex]
\Sigma_{\perp i}= -\!{\rm Im}(A_\perp A_i^*\! +\! {\bar A}_\perp
{{\bar A}_i}^* ),
&&\Sigma_{\| 0}= {\rm Re}(A_\| A_0^*\!-\! {\bar A}_\| {{\bar A}_0}^*
),\nn\\[1.ex]
\rho_{\perp i}\!=\! {\rm Re}\!\Bigl(\frac{q}{p} \!\bigl[A_\perp^*
{\bar A}_i\! +\! A_i^* {\bar A}_\perp\bigr]\Bigr),
&&\rho_{\perp \perp}\!=\! {\rm Im}\Bigl(\frac{q}{p}\, A_\perp^*
{\bar A}_\perp\Bigr),\nn\\[1.ex]
\rho_{\| 0}\!=\! -{\rm Im}\!\Bigl(\frac{q}{p}[A_\|^* {\bar A}_0\! +
\!A_0^* {\bar A}_\| ]\Bigr),
&&\rho_{ii}\!=\! -{\rm Im}\!\Bigl(\frac{q}{p} A_i^* {\bar A}_i\Bigr),
\vspace*{-0.2in}
  \label{eq:obs}
\end{eqnarray}
}
where $i=\{0,\|\}$. In the above, $q/p = \exp({-2\,i\beta^{mix}})$,
where $\beta^{mix}$ is the weak phase describing $\bd$--$\bdbar$
mixing. Note that $\beta^{mix}$ may include NP effects (in the SM,
$\beta^{mix} = \beta$). Note also that the signs of the various $\rho$
terms depend on the CP-parity of the various helicity states. We have
chosen the sign of $\rho_{00}$ and $\rho_{\|\|}$ to be $-1$, which
corresponds to the final state $J/\psi K^*$.

The 18 observables given above can be written in terms of 13
theoretical parameters: three $a_\lambda$'s, three $b_\lambda$'s,
$\beta^{mix}$, $\phi$, and five strong phase differences defined by
$\delta_\lambda \equiv \delta_\lambda^b - \delta_\lambda^a$, $\Delta_i
\equiv \delta_\perp^a - \delta_i^a$. The explicit expressions for the
observables are as follows:
\bea
\Lambda_{\lambda\lambda} & = & a_\lambda^2 + b_\lambda^2 + 2 a_\lambda
b_\lambda \cos\delta_\lambda \cos\phi ~, \nn\\
\Sigma_{\lambda\lambda} & = & - 2 a_\lambda b_\lambda
\sin\delta_\lambda \sin\phi ~, \nn\\
\Lambda_{\perp i} & = & 2 \left[ a_\perp b_i \cos(\Delta_i - \delta_i)
- a_i b_\perp \cos(\Delta_i + \delta_\perp) \right] \sin\phi ~,\nn\\
\Lambda_{\| 0} & = & 2 \left[ a_\| a_0 \cos(\Delta_0 - \Delta_\|) +
a_\| b_0 \cos(\Delta_0 - \Delta_\| - \delta_0) \cos\phi \right. \nn\\
& & ~~ \left. + a_0 b_\| \cos(\Delta_0 - \Delta_\| + \delta_\|)
\cos\phi + b_\| b_0 \cos(\Delta_0 - \Delta_\| + \delta_\| - \delta_0)
\right] ~,\nn\\
\Sigma_{\perp i} & = & -2 \left[ a_\perp a_i \sin \Delta_i + a_\perp
b_i \sin(\Delta_i - \delta_i) \cos\phi 
+ a_i b_\perp \sin(\Delta_i + \delta_\perp) \cos\phi \right. \nn\\
& & ~~ \left. + b_\perp b_i \sin (\Delta_i + \delta_\perp - \delta_i)
\right] ~, \nn\\
\Sigma_{\| 0} & = & 2 \left[ a_\| b_0 \sin(\Delta_0 - \Delta_\| -
\delta_0) - a_0 b_\| \sin(\Delta_0 - \Delta_\| + \delta_\|) \right]
\sin\phi ~, \nn\\
\rho_{ii} & = & a_i^2 \sin 2\beta^{mix} + 2 a_i b_i \cos\delta_i \sin(2
\beta^{mix} + \phi) + b_i^2 \sin(2\beta^{mix} + 2 \phi) ~, \nn\\
\rho_{\perp\perp} & = & - a_\perp^2 \sin 2\beta^{mix} - 2 a_\perp b_\perp
\cos\delta_\perp \sin(2 \beta^{mix} + \phi) - b_\perp^2 \sin(2\beta^{mix} + 2
\phi) ~, \nn\\
\rho_{\perp i} & = & 2 \left[ a_i a_\perp \cos \Delta_i \cos
2\beta^{mix} + a_\perp b_i \cos(\Delta_i - \delta_i) \cos(2
\beta^{mix} + \phi) \right. \nn\\
& & ~~ + a_i b_\perp \cos(\Delta_i + \delta_\perp) \cos(2 \beta^{mix}
+ \phi) \nn\\
& & ~~ \left. + b_i b_\perp \cos(\Delta_i + \delta_\perp - \delta_i)
\cos(2\beta^{mix} + 2\phi) \right] ~,\nn\\
\rho_{\| 0} & = & 2 \left[ a_0 a_\| \cos(\Delta_0 - \Delta_\|) \sin
2\beta^{mix} + a_\| b_0 \cos(\Delta_0 -\Delta_\| - \delta_0) \sin(2
\beta^{mix} + \phi) \right. \nn\\
& & ~~ + a_0 b_\| \cos(\Delta_0 - \Delta_\| + \delta_\|) \sin(2
\beta^{mix} \nn\\
& & ~~ \left. + \phi)+ b_0 b_\| \cos(\Delta_0 - \Delta_\| + \delta_\| -
\delta_0) \sin(2\beta^{mix} + 2\phi) \right] ~.
\label{observables}
\end{eqnarray}
It is straightforward to show that one cannot extract $\beta^{mix}$.
There are a total of six amplitudes describing $B \to V_1 V_2$ and
${\bar B} \to {\overline{V}}_1 {\overline{V}}_2)$ decays
[Eq.~(\ref{amps})]. Thus, at best one can measure the magnitudes and
relative phases of these six amplitudes, giving 11 measurements. Since
the number of meaurements (11) is fewer than the number of theoretical
parameters (13), one cannot obtain any of the theoretical unknowns
purely in terms of observables. In particular, it is impossible to
extract $\beta^{mix}$ cleanly.

In the absence of NP, $b_\lambda = 0$. The number of parameters is
then reduced from 13 to 6: three $a_\lambda$'s, two strong phase
differences ($\Delta_i$), and $\beta^{mix}$. All of these can be
determined cleanly in terms of observables. Because we have 18
observables, but only 6 theoretical parameters, there are 12 relations
that must exist among the observables in the absence of NP. (Of
course, only five of these are independent.) The 12 relations are:
\begin{eqnarray}
&& \Sigma_{\lambda\lambda}= \Lambda_{\perp i}= \Sigma_{\|
  0}=0 \nn\\
&& \frac{\rho_{ii}}{\Lambda_{ii}} =
-\frac{\rho_{\perp\perp}}{\Lambda_{\perp\perp}} =
\frac{\rho_{\|0}}{\Lambda_{\| 0}}\nn \\
&& 
\Lambda_{\|0}=\frac{1}{2\Lambda_{\perp\perp}}\Bigl[
  \frac{\Lambda_{\lambda\lambda}^2\rho_{\perp 0}
  \rho_{\perp\|}+\Sigma_{\perp 0}\Sigma_{\perp
  \|}(\Lambda_{\lambda\lambda}^2
  -\rho_{\lambda\lambda}^2)}
 {\Lambda_{\lambda\lambda}^2-\rho_{\lambda\lambda}^2}\Bigr]\nn
  \\
&& 
  \frac{\rho_{\perp i}^2}{4\Lambda_{\perp\perp}\Lambda_{i
      i}-\Sigma_{\perp i}^2}=\frac{\Lambda_{\perp\perp}^2
    -\rho_{\perp\perp}^2}{\Lambda_{\perp\perp}^2}~.
  \label{eq:no_np}
\end{eqnarray}
The key point is the following: {\em the violation of any of the above
relations will be a smoking-gun signal of NP.} We therefore see that
the angular analysis of $B\to V_1 V_2$ decays provides numerous tests
for the presence of new physics.

The observable $\Lambda_{\perp i}$ deserves special attention
\cite{DattaLondon}. {}From Eq.~(\ref{observables}), one sees that even
if the strong phase differences vanish, $\Lambda_{\perp i}$ is nonzero
in the presence of new physics ($\phi\ne 0$), in stark contrast to the
direct CP asymmetries (proportional to $\Sigma_{\lambda\lambda}$).
This is due to the fact that the $\perp$ helicity is CP-odd, while the
$0$ and $\|$ helicities are CP-even. Thus, $\perp$--$0$ and
$\perp$--$\|$ interferences include an additional factor of `$i$' in
the full decay amplitudes [Eq.~(\ref{fullamps})], which leads to the
cosine dependence on the strong phases.

Now, although the reconstruction of the full $\bd(t)$ and $\bdbar(t)$
decay rates in Eq.~(\ref{decayrates}) requires both tagging and
time-dependent measurements, the $\Lambda_{\lambda\sigma}$ terms
remain even if the two rates for $\bd(t)$ and $\bdbar(t)$ decays are
added together. Note also that these terms are time-independent.
Therefore, {\it no tagging or time-dependent measurements are needed
to extract $\Lambda_{\perp i}$}! It is only necessary to perform an
angular analysis of the final state $V_1 V_2$. Thus, this measurement
can even be made at a symmetric $B$-factory.

The decays of charged $B$ mesons to vector-vector final states are
even simpler to analyze since no mixing is involved.
One can in principle combine charged and neutral $B$
decays to increase the sensitivity to new physics. For example, for $B
\to J/\psi K^*$ decays, one simply performs an angular analysis on all
decays in which a $J/\psi$ is produced accompanied by a charged or
neutral $K^*$. A nonzero value of $\Lambda_{\perp i}$ would be a clear
signal for new physics \cite{FPCP}.

The decays of both charged and neutral $B$ mesons to $D_s^* D^*$ can
be analyzed similarly. Because these modes are dominated by a single
decay amplitude in the SM, no direct CP violation is expected. And
since this is not a final state to which both $\bd$ and $\bdbar$ can
decay, no indirect CP violation is present either. An angular analysis
of these decays would therefore be interesting -- if any CP-violating
signal were found, such as a nonzero value of $\Lambda_{\perp i}$,
this would again indicate the presence of new physics.

It must be noted that, despite the large number of new-physics
signals, it is still possible for the NP to remain hidden. This
happens if a singular situation is realized. If the three strong phase
differences $\delta_\lambda$ vanish, and the ratio $r_\lambda \equiv
b_\lambda/a_\lambda$ is the same for all helicities, i.e. $r_0 = r_\|
= r_\perp$, then it is easy to show that the relations in
Eq.~(\ref{eq:no_np}) are all satisfied. Thus, if the NP happens to
respect these very special conditions, the angular analysis of $B\to
V_1 V_2$ would show no signal for NP even if it is present, and the
measured value of $\beta$ would not correspond to the actual phase of
$\bd$--$\bdbar$ mixing.

Now, suppose that some signal of new physics is found, indicating that
$\beta^{meas}$ is not equal to $\beta^{mix}$. As we have argued
earlier, in the presence of new physics one cannot extract the true
value of $\beta^{mix}$. However, as we will describe below, the
angular analysis does allow one to constrain the value of the
difference $|\beta^{\mathit{meas}} - \beta^{mix}|$, as well as the
size of the NP amplitudes $b^2_\lambda$. Naively, one would not think
it possible to obtain any constraints on the NP parameters. After all,
we have 11 measurements, but 13 theoretical unknown parameters.
However, because the equations are nonlinear, such constraints are
possible. Below, we list some of these constraints; their full
derivation will be presented elsewhere \cite{LSSnew}.

In the constraints, we will make use of the following quantities. For
the vector-vector final state, the analogue of the usual direct CP
asymmetry $a^{CP}_{dir}$ is $a_{\lambda}^{dir} \equiv
\Sigma_{\lambda\lambda}/\Lambda_{\lambda\lambda}$, which is
helicity-dependent. For convenience, we define the related quantity
$y_\lambda = \sqrt{1 - \Sigma_{\lambda\lambda}^2/
\Lambda_{\lambda\lambda}^2}$. The measured value of $\sin 2\beta$ can
also depend on the helicity of the final state:
$\rho_{\lambda\lambda}$ can be recast in terms of a measured weak
phase $2\beta^{\mathit{meas}}_{\lambda}$, defined as
\begin{equation}
\sin\, 2\,\beta^{\mathit{meas}}_{\lambda}=\frac{\pm
\rho_{\lambda\lambda}}{\sqrt{\Lambda^2_{\lambda\lambda}-
\Sigma^2_{\lambda\lambda}}}~,
\end{equation}
where the $+$ $(-)$ sign corresponds to $\lambda=0,\|$ ($ \perp$). In
terms of these quantities, the size of NP amplitudes $b_\lambda^2$ may
be expressed as
\beq
2\,b_\lambda^2\,\sin^2\phi = \Lambda_{\lambda\lambda}
    \Big(1-y_\lambda\cos(2\beta^{meas}_\lambda-2\beta)\Big) ~.
\label{eq:a-b-2}
\eeq

The form of the constraints depends on which new-physics signals are
observed; we give a partial list below. For example, suppose that
direct CP violation is observed in a particular helicity state. In
this case a lower bound on the corresponding NP amplitude can be
obtained by minimizing $b^2_\lambda$ [Eq.~(\ref{eq:a-b-2})] with
respect to $\beta$ and $\phi$:
\beq
b^2_\lambda \ge {1\over 2} \Lambda_{\lambda\lambda} \left[ 1 -
y_\lambda \right].
\eeq
On the other hand, suppose that the new-physics signal is
$\beta^{\mathit{meas}}_i \ne \beta^{\mathit{meas}}_j$. Defining
$2\omega \equiv 2\beta^{\mathit meas}_j-2\beta^{\mathit meas}_i$ and
$\eta_\lambda \equiv 2 ( \beta^{\mathit{meas}}_\lambda - \beta^{mix}
)$, the minimization of $(b_i^2 \mp b_j^2)$ with respect to $\eta_i$
and $\phi$ yields
\beq
(b_i^2 \mp b_j^2) \ge \frac{\Lambda_{ii} \mp \Lambda_{jj}}{2} - \frac{
\left\vert y_i \Lambda_{ii} \mp y_j \Lambda_{jj} e^{2 i \omega}
\right\vert }{2} ~,
\label{eq:bsq-omega-bounds}
\eeq
where $\Lambda_{ii} > \Lambda_{jj}$ is assumed. If there is no direct
CP violation ($\Sigma_{\lambda\lambda} = 0$), but $\Lambda_{\perp i}$
is nonzero, one has
\beq
2 (b_i^2 \mp b_\perp^2) \geq \Lambda_{ii} \mp \Lambda_{\perp\perp} -
\sqrt{ \left( \Lambda_{ii} \mp \Lambda_{\perp\perp}\right)^2 \pm
\Lambda_{\perp i}^2} ~,
\label{eq:bsq-Lambda-bounds}
\eeq
where $\beta^{meas}_\perp$ was eliminated using the expression for
$\Lambda_{\perp i}$.

One can also obtain bounds on $|\beta^{\mathit{meas}}_\lambda -
\beta^{mix}|$, though this requires the nonzero measurement of
observables involving the interference of different helicities. For
example, if $\Lambda_{\perp i}$ is nonzero and
$\Sigma_{\lambda\lambda} = 0$, we find
\[
\Lambda_{ii} \cos\eta_i + \Lambda_{\perp\perp} \cos(\eta_\perp - 2
\eta_i) \le \sqrt{ \left( \Lambda_{ii} + \Lambda_{\perp\perp}
\right)^2 - \Lambda_{\perp i}^2}~,
\]
\null\vskip-12truemm
\beq
\Lambda_{ii} \cos\eta_i - \Lambda_{\perp\perp} \cos \eta_\perp
\le \sqrt{ \left( \Lambda_{ii} - \Lambda_{\perp\perp}
\right)^2 + \Lambda_{\perp i}^2}~.
\eeq
If $\Lambda_{\perp i} \ne 0$, one cannot have $\eta_i = \eta_\perp =
0$. These constraints therefore place a lower bound on
$|\beta^{\mathit{meas}}_i - \beta^{mix}|$ and/or
$|\beta^{\mathit{meas}}_\perp - \beta^{mix}|$.

A-priori, one does not know which of the above constraints is
strongest -- this depends on the actual values of the observables. Of
course, in practice, one will simply perform a fit to obtain the best
lower bounds on these NP parameters \cite{LSSnew}. However, it is
interesting to see that constraints can be obtained analytically.

As a specific application, we have noted the apparent discrepancy in
the value of $\sin 2\beta$ as obtained from measurements of $\bd(t)\to
J/\psi \ks$ and $\bd(t) \to \phi \ks$ \cite{phiKs}. In this case, the
angular analyses of $\bd(t)\to J/\psi K^*$ and $\bd(t) \to \phi K^*$
would allow one to determine if new physics is indeed present. If NP
is confirmed, this analysis would allow one to put constraints on the
NP parameters.

Note that this analysis can also be applied within the SM to decays
such as $\bd(t) \to D^{*+} D^{*-}$. These decays have both a tree and
a penguin contribution, so that $\beta^{mix}$ cannot be extracted
cleanly. Assuming no new physics, the above analysis allows one to
obtain lower bounds on the ratio of penguin to tree amplitudes, as
well as on $|\beta^{\mathit{meas}}_\lambda - \beta^{mix}|$. This can
serve as a crosscheck on other measurements of $\beta^{mix}$, as well
as on model calculations of the hadronic amplitudes.

Finally, it is worthwhile to examine the feasibility of this method.
The present data at $B$-factories can already be used to perform
time-independent angular analyses of $B\to V_1 V_2$ decays. In fact,
BaBar has measured a nonzero value of $\Lambda_{\perp i}$ (a
CP-violating triple-product correlation) in $B \to \phi K^*$ at
$1.7\sigma$ \cite{TPsignal}. This is a potential hint of new physics.
On the other hand, time-dependent angular analyses will take
considerably more time to carry out. Thus, it may be several more
years before we have new-physics signals which rely on time-dependent
measurements.

To sum up: in the standard model (SM), the cleanest extraction of the
CP angles comes from neutral $B$ decays that are dominated by a single
decay amplitude. If there happens to be a new-physics (NP)
contribution to the decay amplitude, with a different weak phase, this
could seriously affect the cleanliness of the measurement. There is
already a hint of such NP, as indicated by the discrepancy between the
value of $\beta$ extracted from $\bd(t)\to J/\psi \ks$ and that
obtained from $\bd(t) \to \phi \ks$. However, it is important to
confirm this through independent direct tests, and to attempt to
obtain information about the NP amplitude, if possible.

In this paper, we have shown that this type of new physics can be
probed by performing an angular analysis of the related $B \to V_1
V_2$ decay modes. There are numerous relations that are violated in
the presence of NP, and several of these remain nonzero even if the
strong phase difference between the SM and NP amplitudes vanishes. The
most incisive test is a measurement of $\Lambda_{\perp i} \ne 0$. To
obtain this observable, neither tagging nor time-dependent
measurements is necessary -- one can combine all neutral and charged
$B$ decays.

Furthermore, should a signal for new physics be found, one can place a
lower bound on the difference $|\beta^{\mathit{meas}} - \beta^{mix}|$,
as well as on the size of the NP amplitudes. By applying this analysis
to the decays $\bd(t)\to J/\psi K^*$ and $\bd(t) \to \phi K^*$, one
can confirm the presence of the new physics that is hinted at in the
measurements of $\bd(t)\to J/\psi \ks$ and $\bd(t) \to \phi \ks$
\cite{phiKs}. It can even be applied within the SM to decays such as
$\bd(t) \to D^{*+} D^{*-}$, which receive both tree and penguin
contributions.

\bigskip
\noindent
{\bf Acknowledgements}: N.S. and R.S. thank D.L. for the hospitality
of the Universit\'e de Montr\'eal, where part of this work was
done. The work of D.L. was financially supported by NSERC of
Canada. The work of N.S. was supported by the Department of Science
and Technology, India.


\end{document}